\documentclass[11pt]{article}
\usepackage{bbold}
\usepackage{geometry}
\usepackage{amsmath}

\usepackage{amsfonts}
\usepackage{amssymb}
\usepackage{graphicx}
\usepackage{caption}
\usepackage{subcaption}
\usepackage{hyperref}
\usepackage{float}
\usepackage{braket}
\usepackage{calrsfs}
\usepackage{mathrsfs}
\usepackage{fancyhdr}
\usepackage{multicol}
\usepackage[font=footnotesize]{caption}

\makeatletter
\def\maketag@@@#1{\hbox{\m@th\normalfont\normalsize#1}}
\makeatother

\title{\vspace{-1.7cm}\textbf{Eternal life of entropy in non-Hermitian quantum systems}}
\author{\textbf{Andreas Fring and Thomas Frith}\\\textit{Department of Mathematics, City, University of London,}\\
	\textit{Northampton Square, London EC1V 0HB, UK}\\
	\textit{E-mail: a.fring@city.ac.uk, thomas.frith@city.ac.uk}\vspace{-0.5cm}}
\date{}
\numberwithin{equation}{section}
\numberwithin{equation}{section}

\begin{document}

\maketitle 
\thispagestyle{fancy}

ABSTRACT: We find a new effect for the behaviour of Von Neumann entropy. For this we derive the framework for describing Von Neumann entropy in non-Hermitian quantum systems and then apply it to a simple interacting $PT$ symmetric bosonic system. We show that our model is well defined even in the $PT$ broken regime with the introduction of a time-dependent metric and that it displays three distinct behaviours relating to the $PT$ symmetry of the original time-independent Hamiltonian. When the symmetry is unbroken, the entropy undergoes rapid decay to zero (so-called "sudden death") with a subsequent revival. At the exceptional point it decays asymptotically to zero and when the symmetry is spontaneously broken it decays asymptotically to a finite constant value ("eternal life").\vspace{0cm}

\begin{multicols}{2}

\section{Introduction}

The information contained within a quantum system is of great importance for various practical implementations of quantum mechanics, most importantly for the development of quantum computers, e.g.  \cite{nielsen2002quantum,bennett2000quantum,raussendorf2001one,steane1998quantum}. In order to understand the quantum information, one must find a way of measuring the entanglement of a state. Entanglement is a defining feature of quantum mechanics that distinguishes it from classical mechanics and there has been much work in recent years into the evolution of entanglement with time, particularly the observation of the abrupt decay of entangled states, coined as "sudden death" \cite{yu2009sudden,yonacc2006sudden}. The decoherence of entanglement \cite{unruh1995maintaining,palma1996quantum} is a problem for the operation of quantum computers and so understanding the mechanism behind this is an important contribution to the development of future machines. One particular measurement of entanglement and quantum information is the Von Neumann entropy. This is well-understood in the standard quantum mechanical setting, however to date there has only been a small amount of work done concerning the proper treatment of entropy in non-Hermitian, parity-time ($PT$) symmetric systems \cite{scolarici2006time,croke2015pt,kawabata2017information,jones2010quantum}. These differ from open quantum systems as the energy eigenvalues are real or appear as complex conjugate pairs and do not describe decay. 

Non-Hermitian, parity-time ($PT$) symmetric quantum mechanics was first popularised when it was shown that non-Hermitian systems with unbroken $PT$ symmetry had real eigenvalues and unitary time evolution  \cite{bender1998real,bender2007making,mostafazadeh2010pseudo,moiseyev2011non,scholtz1992quasi}. This is possible due to the existence of a non-trivial metric operator and much work has been done on constructing metrics for time-independent systems, e.g. \cite{znojil2006construction,siegl2012metric,castro2009spin,jones2007scattering,musumbu2006choice,mostafazadeh2006metric}. More recently this has extended to time-dependent systems, e.g.  \cite{mostafazadeh2007time,fring2016non,fring2016unitary,maamache2017pseudo,znojil2008time,de2006time,mostafazadeh2018energy}. Of particular interest are non-Hermitian systems with spontaneously broken $PT$ symmetry. These systems possess an exceptional point above which the $PT$ symmetry is broken. In this regime the system exhibits complex energy eigenvalues, becoming ill-defined and is therefore ordinarily discarded as non-physical and useless. However, it has been shown \cite{ExactSols,FRING20172318,HigherSpin,fring2018tdm} that when a time-dependence is introduced into the central equations it is possible to make sense of the broken regime via a time-dependent metric. This allows for the definition of a Hilbert space and therefore a well-defined inner product. This will be central to our analysis in non-Hermitian systems as we will be showing how the evolution of entropy changes significantly as we vary the system parameters through the exceptional point. 

We will first set up the framework for analysing the Von Neumann entropy in non-Hermitian systems and then we will apply it to a simple model consisting of a bosonic system coupled to a bath. 

\vspace{-0.4cm}

\section{Entanglement Von Neumann Entropy}\label{Theory}

In order to make calculations of the quantum entropy for non-Hermitian systems, we must first introduce some new quantities when compared to the Hermitian case. The density matrix for Hermitian systems is defined as an Hermitian operator describing the statistical ensemble of states

\begin{equation}
\varrho_h=\sum_{i}p_i\ket{\phi_i}\bra{\phi_i},
\end{equation}
\newpage

\rhead{}
\lhead{Eternal life of entropy in non-Hermitian quantum systems} 

\noindent 
where the subscript $h$ indicates it relates to an Hermitian system. $\ket{\phi_i}$ are general pure states, and $p_i$ is the probability that the system is in the pure state $\ket{\phi_i}$, with  $0\leq p_i\leq 1$ and $\sum_ip_i=1$. Therefore $\varrho_h$ represents a mix of pure states (a mixed state). If the system is comprised of subsystems $A$ and $B$ one can define the reduced density operator of these subsystems as the partial trace over the opposing subsystem's Hilbert space

\begin{equation}
\varrho_{h,A}=Tr_B\left[\varrho_h\right]=\sum_{i}\bra{n_{i,B}}\varrho_h\ket{n_{i,B}},
\end{equation}
\begin{equation}
\varrho_{h,B}=Tr_A\left[\varrho_h\right]=\sum_{i}\bra{n_{i,A}}\varrho_h\ket{n_{i,A}},
\end{equation}
where $\ket{n_{i,A}}$ and $\ket{n_{i,B}}$ are the eigenstates of the subsystems $A$ and $B$, respectively. In this way one can isolate the density matrix for each subsystem and perform entropic analysis on them individually. We now want to find the relationship between the $\varrho_h$ and $\varrho_H$, where the subscript $H$ indicates a non-Hermitian system. The clearest starting point is the Von Neumann equation which governs the time evolution of the density matrix. For the Hermitian system it is

\begin{equation}\label{VNHermitian}
i\partial_t\varrho_h=\left[h,\varrho_h\right],
\end{equation}
where $h$ is the Hermitian Hamiltonian. We now wish to find the equivalent relation in the non-Hermitian setting. In order to do this we substitute the time-dependent Dyson equation \cite{FRING20172318,HigherDims}

\begin{equation}\label{TDDE}
h=\eta H\eta^{-1}+i\partial_t\eta \eta^{-1},
\end{equation}
into the Von Neumann equation. $\eta$ is the Dyson operator and forms the metric $\rho=\eta^\dagger\eta$. After some manipulation, substituting equation (\ref{TDDE}) into (\ref{VNHermitian})  results in the following equation

\begin{equation}
i\partial_t\varrho_H=\left[H,\varrho_H\right],
\end{equation}
when assuming that the density matrix in the Hermitian system is related to that of the non-Hermitian system via a similarity transformation

\begin{equation}\label{SimTrans}
\varrho_h=\eta\varrho_H\eta^{-1}.
\end{equation}

Recalling that $\ket{\phi}=\eta\ket{\psi}$, this leads us to the definition of the density matrix $\varrho_H$ for non-Hermitian systems

\begin{equation}
\varrho_H=\sum_{i}p_i\ket{\psi_i}\bra{\psi_i}\rho,
\end{equation}
%This must be included as we are performing a spectral decomposition with respect to the inner product and the non-Hermitian inner product contains the metric. 
where $\ket{\psi_i}$ are general pure states for the non-Hermitian system. Notice that $\varrho_H$ is an Hermitian operator in the Hilbert space related to the metric $\bra{\cdot}\rho\ket{\cdot}$. These results match those from \cite{scolarici2006time}. Having defined the density matrix for non-Hermitian systems and found the relation to Hermitian systems we can now consider the entropy. For the total system, the Von Neumann entropy is defined as

\begin{equation}
S_h=-tr\left[\rho_h\ln\rho_h\right].
\end{equation}
This can also be expressed as a sum of the eigenvalues $\lambda_i$ of the density matrix $\rho_h$ as it is an Hermitian operator

\begin{equation}
S_h=-\sum_{i}\lambda_i\ln\lambda_i.
\end{equation}
As the density matrix for the Hermitian and non-Hermitian systems are related by a similarity transform, they share the same eigenvalues, therefore

\begin{equation}
S_H=S_h.
\end{equation}
Is is important to recall, however, that this relation only holds true for the existence of a well-defined Dyson operator $\eta$. Without this, we are unable to form the relation (\ref{SimTrans}). For closed systems, the Von Neumann entropy is constant with time. However, we wish to consider the entropy for particular subsystems and for this we must consider the partial trace of the density matrix. In this setting the entropy for subsystem $A$ becomes

\begin{equation}
S_{h,A}=-tr\left[\rho_{h,A}\ln\rho_{h,A}\right]=-\sum_{i}\lambda_{i,A}\ln\lambda_{i,A},
\end{equation}
where once again the entropy of the Hermitian subsystem is equal to that of the non-Hermitian subsystem $S_{h,A}=S_{H,A}$ with the existence of $\eta$. The entropy of a particular subsystem is not confined to be constant and we show that it exhibits some very interesting properties when evolved in time.

%This is in contrast to \cite{sergi2016quantum}.

\section{System bath coupled model}

We now consider a time-independent non-Hermitian Hamiltonian consisting of a bosonic system coupled to a bath of $N$ bosonic systems. The Hamiltonian takes the form

\begin{equation}\label{NHHamiltonian}
\begin{split}
\begin{aligned}
H=&\nu a^\dagger a+\nu\sum_{n=1}^{N}q^\dagger_nq_n+\left(g+\kappa\right)a^\dagger\sum_{n=1}^{N}q_n\\
+&\left(g-\kappa\right)a\sum_{n=1}^{N}q^\dagger_n,
\end{aligned}
\end{split}
\end{equation}
with $\nu$, $g$ and $\kappa$ being real time-independent parameters.

\subsection{$PT$ symmetry}
The Hamiltonian (\ref{NHHamiltonian}) is PT symmetric under the anti-linear transformation 

\begin{equation}
\begin{split}
\begin{aligned}
PT: \quad i&\rightarrow -i, \quad a\rightarrow-a, \quad a^\dagger\rightarrow-a^\dagger, \\ q_n&\rightarrow-q_n, \quad q_n^\dagger\rightarrow-q_n^\dagger,
\end{aligned}
\end{split}
\end{equation}
as it commutes with the $PT$ operator for all values of $\nu$, $g$ and $\kappa$

\begin{equation}
\left[PT,H\right]=0,
\end{equation}
The energy eigenvalues are 

\begin{equation}
E_{m,N}^\pm=m\left(\nu\pm\sqrt{N}\sqrt{g^2-\kappa^2}\right).
\end{equation}
In order to ensure boundedness from below the system must have $\nu>\sqrt{N}\sqrt{g^2-\kappa^2}$. Note that there is an exceptional point at $g=\kappa$ and when $\kappa>g$ this system is in the broken $PT$ regime. This is clear when studying the first excited state ($m=1$) expanded in terms of creation operators acting on a tensor product of Fock states. The general state consists of one Fock state for the system of $a$ and $a^\dagger$ bosonic operators and $N$ Fock states for the bath of $q_i$ and $q_i^\dagger$ bosonic operators

\begin{equation}
\ket{\phi}=\ket{n_a}\otimes\ket{n_{q_1}}\otimes\ket{n_{q_2}}....=\ket{n_a}\bigotimes_{i=1}^{N}\ket{n_{q_i}}.
\end{equation}
When considering the first excited state, we will be dealing with very few non-zero states, and as such we can make some simplifications to the notation. If all the states in the $q$ bath are in the ground state we will represent this with $\ket{\boldmath{0}_q}$. Similarly, if the $i$th state in the $q$ bath is in the first excited state with the rest in the ground state, we will represent this with a $\ket{\boldmath{1}_{i}}$

\begin{equation}
\begin{split}
\begin{aligned}
\ket{\boldmath{0}_q}&=\bigotimes_{i=1}^{N}\ket{0_{q_i}}, \\ \ket{\boldmath{1}_{i}}&=\left[\bigotimes_{j=1}^{i-1}\ket{0_{q_j}}\right]\otimes\ket{1_{q_i}}\otimes\left[\bigotimes_{k=i+1}^{N}\ket{0_{q_k}}\right].
\end{aligned}
\end{split}
\end{equation}
We can now write down the first excited state, 
\begin{equation}
\begin{aligned}
\begin{split}
\ket{\psi_{1,N}^\pm}=&\sqrt{\frac{g+\kappa}{2g}}\ket{1_a}\otimes\ket{\boldmath{0}_q}\pm\sqrt{\frac{g-k}{2gN}}\ket{0_a}\otimes\sum_{i=1}^{N}\ket{ 1_{i}}\\
=&\sqrt{\frac{g+\kappa}{2g}}\ket{1_a\boldmath{0}_q}\pm\sqrt{\frac{g-k}{2gN}}\sum_{i=1}^{N}\ket{0_a 1_{i}}\\
=&\sqrt{\frac{g+\kappa}{2g}}a^\dagger\ket{0_a\boldmath{0}_q}\pm\sqrt{\frac{g-k}{2gN}}\sum_{i=1}^{N}q_i^\dagger\ket{0_a \boldmath{0}_{q}}.
\end{split}
\end{aligned}
\end{equation}
%We can expand this in terms of creation operators
%
%\begin{equation}
%\ket{\psi_{1,N}}=
%\end{equation}
In order for the $PT$ symmetry to remain unbroken, the wavefunction must also remain unchanged up to a phase factor when acted on by the $PT$ operator

\begin{equation}
PT\ket{\psi_{1,N}^\pm}= e^{i\phi}\ket{\psi_{1,N}^\pm}.
\end{equation}
However, the wavefunctions are only eigenfunctions of the $PT$ operator when $\kappa<g$
%As an example, when $N=2$ the first excited state is

%\begin{equation}
%\ket{\psi_{1,2}}=\sqrt{\frac{g+\kappa}{2g}}\ket{100}+\frac{\sqrt{g-k}}{\sqrt{4g}}\left(\ket{0 10}+\ket{0 01}\right).
%\end{equation}

\begin{equation}
PT\ket{\psi_{1,N}^\pm}=-\ket{\psi_{1,N}^\pm}.
\end{equation}
When $\kappa>g$, the wavefunctions are no longer eigenfunctions of the $PT$ operator,

\begin{equation}
PT\ket{\psi_{1,N}^\pm}\neq e^{i\phi}\ket{\psi_{1,N}^\pm}.
\end{equation}
Therefore we need to employ time-dependent analysis in order to make sense of the broken regime. To do this we first must solve the time-dependent Dyson equation. 

\subsection{Solving the time-dependent Dyson equation}

We wish to find the time-dependent metric $\rho\left(t\right)$ that allows us to perform entropic analysis on our model (\ref{NHHamiltonian}). In order to do this we must find the Dyson operator $\eta\left(t\right)$ and the equivalent time-dependent Hermitian system $h\left(t\right)$. The model (\ref{NHHamiltonian}) is in fact part of a larger family of Hamiltonians belonging to the closed algebra with Hermitian generators:

\begin{equation}
\begin{aligned}
\begin{split}
N_A&=a^\dagger a, \qquad N_Q=\sum_{n=1}^{N}q^\dagger_nq_n, \\ N_{AQ}&=N_A-\frac{1}{N}N_Q-\frac{1}{N}\sum_{n\neq m}q^\dagger_nq_m \\ A_x&=\frac{1}{\sqrt{N}}\left(a^\dagger\sum_{n=1}^{N}q_n+a\sum_{n=1}^{N}q^\dagger_n\right), \\ A_y&=\frac{i}{\sqrt{N}}\left(a^\dagger\sum_{n=1}^{N}q_n-a\sum_{n=1}^{N}q^\dagger_n\right).
\end{split}
\end{aligned}
\end{equation}
The commutation relations are

\begin{eqnarray}
\begin{aligned}
\begin{split}
\left[N_A,N_Q\right]&=0, \quad\qquad \left[N_A,N_{AQ}\right]=0, \\
\left[N_A,A_x\right]&=-iA_y, \quad\quad \left[N_A,A_y\right]=iA_y,\\
\left[N_Q,A_x\right]&=iA_y, \quad\quad\;\;\; \left[N_Q,A_y\right]=-iA_x, \\ 
\left[N_{AQ},A_x\right]&=-2iA_y, \quad \left[N_{AQ},A_y\right]=2iA_x.
\end{split}
\end{aligned}
\end{eqnarray}
In terms of this algebra, our original Hamiltonian (\ref{NHHamiltonian}) can be written as

\begin{equation}
H=\nu N_A+\nu N_Q+ \sqrt{N}gA_x-i\sqrt{N}\kappa A_y.
\end{equation}
We are now in a position to begin solving the time-dependent Dyson equation (\ref{TDDE}). For this we make the ansatz

\begin{equation}\label{eta}
\eta\left(t\right)=e^{\beta\left(t\right)A_y}e^{\alpha\left(t\right)N_{AQ}},
\end{equation}
and use the Baker-Campbell-Hausdourff formula to expand the Dyson equation (\ref{TDDE}) in terms of generators. In order to make the resulting Hamiltonian Hermitian, we must solve two coupled differential equations to eliminate the non-Hermitian terms.

\begin{equation}
\dot{\alpha}=-\tanh\left(2\beta\right)\left[\sqrt{N}g\cosh\left(2\alpha\right)+\sqrt{N}\kappa\sinh\left(2\alpha\right)\right],\label{alphadot}
\end{equation}

\begin{equation}
\hspace{-2.3cm}\dot{\beta}=\sqrt{N}\kappa\cosh\left(2\alpha\right)+\sqrt{N}g\sinh\left(2\alpha\right).\label{betadot}
\end{equation}
Equation (\ref{betadot}) can be solved for $\alpha$,

\begin{equation}\label{alpha}
\tanh \left(2\alpha\right)=\frac{-N g\kappa+\dot{\beta}\sqrt{\dot{\beta}^2+N\left( g^2-\kappa^2\right)}}{Ng^2+\dot{\beta}^2}.
\end{equation}
In principle this could lead to a restriction to the term on the RHS of equation (\ref{alpha}) as $-1<\tanh\left(2\alpha\right)<1$. However as we will see, this restriction is obeyed with the final solutions for $\alpha$ and $\beta$. Substituting (\ref{alpha}) into equation (\ref{alphadot}) gives

\begin{equation}
\ddot{\beta}+2\tanh\left(2\beta\right)\left[Ng^2-N\kappa^2+\dot{\beta}^2\right]=0.
\end{equation}
Now making the substitution $\sinh\left(2\beta\right)=\sigma$, this reverts to an harmonic oscillator equation

\begin{equation}
\ddot{\sigma}+4N\left(g^2-\kappa^2\right)\sigma=0,
\end{equation}
which is solved with the function 

\begin{equation}
\sigma=\frac{c_1}{\sqrt{g^2-\kappa^2}}\sin\left(2\sqrt{N}\sqrt{g^2-\kappa^2}\left(t+c_2\right)\right),
\end{equation}
for all values of $\kappa$, where $c_1$ and $c_2$ are constants of integration. We can now write down expressions for $\alpha$ and $\beta$

\begin{equation}
\hspace{-5.6cm}\tanh\left(2\alpha\right)=\frac{\zeta^2-1}{\zeta^2+1},
\end{equation}

\begin{equation}
\sinh\left(2\beta\right)=\frac{c_1}{\sqrt{g^2-\kappa^2}}\sin\left(2\sqrt{N}\sqrt{g^2-\kappa^2}\left(t+c_2\right)\right),
\end{equation}
where $\zeta$ is of the form

\begin{equation}
\footnotesize
\begin{aligned}
\begin{split}
\zeta=\sqrt{2}\sqrt{\frac{g-\kappa}{g+\kappa}}\left[
\frac{\sqrt{c_1^2+g^2-\kappa^2}+c_1\cos\left(2\sqrt{N}\sqrt{g^2-\kappa^2}\left(t+c_2\right)\right)}{\sqrt{c_1^2+2\left(g^2-\kappa^2\right)-c_1^2\cos\left(4\sqrt{N}\sqrt{g^2-\kappa^2}\left(t+c_2\right)\right)}}\right].
\end{split}
\end{aligned}
\end{equation}
Therefore we have a well-defined solution for $\eta\left(t\right)$ from our original ansatz (\ref{eta}) which results in the following time-dependent Hermitian Hamiltonian

\begin{equation}\label{HHamiltonian}
h\left(t\right)=\nu N_A+\nu N_Q+\mu\left(t\right)Ax,
\end{equation}
where 

\begin{equation}\label{mu}
\mu\left(t\right)=\frac{\left(g^2-\kappa^2\right)\sqrt{N}\sqrt{c_1^2+g^2-\kappa^2}}{c_1^2+2\left(g^2-\kappa^2\right)-c_1^2\cos\left(4\sqrt{N}\sqrt{g^2-\kappa^2}\left(t+c_2\right)\right)}.
\end{equation}
This is real provided $|\frac{c1}{\sqrt{g^2-\kappa^2}}|>1$. The general time-dependent first excited state is 

\begin{equation}\label{1State}
\begin{aligned}
\begin{split}
\ket{\phi\left(t\right)}&=e^{-i\nu t}\left(A\sin\mu_I\left(t\right)+B\cos\mu_I\left(t\right)\right)\ket{1_a\boldmath{0}_q}\\
+&\frac{e^{-i\nu t}}{\sqrt{N}}\left(A\cos\mu_I\left(t\right)-B\sin\mu_I\left(t\right)\right)\sum_{i=1}^{N}\ket{0_a \boldmath{1}_{i}},
\end{split}
\end{aligned}
\end{equation}
with $A^2+B^2=1$ and

\begin{equation}
\small
\begin{aligned}
\begin{split}
\mu_I&\left(t\right)=\int^t\mu\left(s\right)ds=\\
\frac{1}{2}&\arctan\left(\frac{\sqrt{c_1^2+g^2-\kappa^2}\tan\left(2\sqrt{N}\sqrt{g^2-\kappa^2}\left(t+c_2\right)\right)}{\sqrt{g^2-\kappa^2}}\right).
\end{split}
\end{aligned}
\end{equation}
Now we have a full solution for $\eta\left(t\right)$ and therefore $\rho\left(t\right)=\eta\left(t\right)^\dagger\eta\left(t\right)$. This allows us to calculate the entropy for our non-Hermitian system (\ref{NHHamiltonian}). The easiest route to take is to work with the resulting Hermitian system (\ref{HHamiltonian}) as it was shown in section \ref{Theory} that the entropy in both systems is equivalent when $\eta\left(t\right)$ is well-defined. It is important to note that if the $\eta\left(t\right)$ ever becomes ill-defined, then our analysis of the Hermitian system does not correspond to the original non-Hermitian Hamiltonian.

\section{Three types of entropy evolution}

We now calculate the entropy of the system and show how varying the parameters $N$, $g$ and $\kappa$ affect its evolution with time. We prepare our system in an entangled first excited state (\ref{1State}) at time $t=0$, this is equivalent to a single qubit entangled with itself.

\begin{equation}
\ket{\phi\left(0\right)}=\sin\gamma\ket{{1_a \boldmath{0}_q}}+\frac{\cos\gamma}{\sqrt{N}}\sum_{i=1}^N\ket{{0_a\boldmath{1}_{i}}},
\end{equation}
for which we choose $A=\sin\gamma$, $B=\cos\gamma$ and $c_2=0$. Therefore the general state at time $t$ is

\begin{equation}
\begin{aligned}
\begin{split}
\ket{\phi\left(t\right)}&=e^{-i\nu t}\left(\sin\gamma\sin\mu_I\left(t\right)+\cos\gamma\cos\mu_I\left(t\right)\right)\ket{1_a\boldmath{0}_q}\\
+&\frac{e^{-i\nu t}}{\sqrt{N}}\left(\sin\gamma\cos\mu_I\left(t\right)-\cos\gamma\sin\mu_I\left(t\right)\right)\sum_{i=1}^{N}\ket{0_a \boldmath{1}_{i}}.
\end{split}
\end{aligned}
\end{equation}
Now we form the density matrix for the system (a) with a partial trace over the external bosonic bath (q),

\begin{equation}
\begin{split}
\begin{aligned}
\varrho_a\left(t\right)&=Tr_q\left[\varrho_h\left(t\right)\right]=\\
&\left(\begin{array}{cc}
\left(\sin\gamma\sin\mu_I\left(t\right)+\cos\gamma\cos\mu_I\left(t\right)\right)^2 & \hspace{-4.5cm}0\hspace{-4.5cm}\\
\hspace{-4.5cm }0 & \hspace{-4.5cm} \left(\sin\gamma\cos\mu_I\left(t\right)-\cos\gamma\sin\mu_I\left(t\right)\right)^2
\end{array}
\right).
\end{aligned}
\end{split}
\end{equation}
We can now calculate the Von Neumann entropy of the system using this reduced density matrix. First we read off the eigenvalues of $\varrho_a\left(t\right)$ as it is diagonal,

\begin{equation}
\begin{split}
\begin{aligned}
\lambda_1\left(t\right)&=\left(\sin\gamma\sin\mu_I\left(t\right)+\cos\gamma\cos\mu_I\left(t\right)\right)^2,\\
\lambda_2\left(t\right)&=\left(\sin\gamma\cos\mu_I\left(t\right)-\cos\gamma\sin\mu_I\left(t\right)\right)^2,\\
\end{aligned}
\end{split}
\end{equation}
and substitute these into the expression for the entropy

\begin{equation}
S_{h,a}\left(t\right)=S_{H,a}\left(t\right)=-\lambda_1\left(t\right)\log\left[\lambda_1\left(t\right)\right]-\lambda_2\left(t\right)\log\left[\lambda_2\left(t\right)\right].
\end{equation}
With this expression we are free to choose the initial state of our system with a given value of $\gamma$. If the initial state of our system is maximally entangled state with $\gamma=\pi/4$, then we observe how the entanglement entropy evolves with time. This is most applicable to quantum computing as in that context one would like to preserve the entangled state. We will now vary the parameters $N$, $g$ and $\kappa$ to see how they affect the evolution of entropy with time. Of particular interest is the exceptional point $g=\kappa$ where the non-Hermitian system enters the broken $PT$ regime in the time-independent setting. It is in this area that the evolution we see differs from the standard evolution of entropy in Hermitan quantum mechanics. 

\begin{figure}[H]
	\centering
	\includegraphics[scale=0.448]{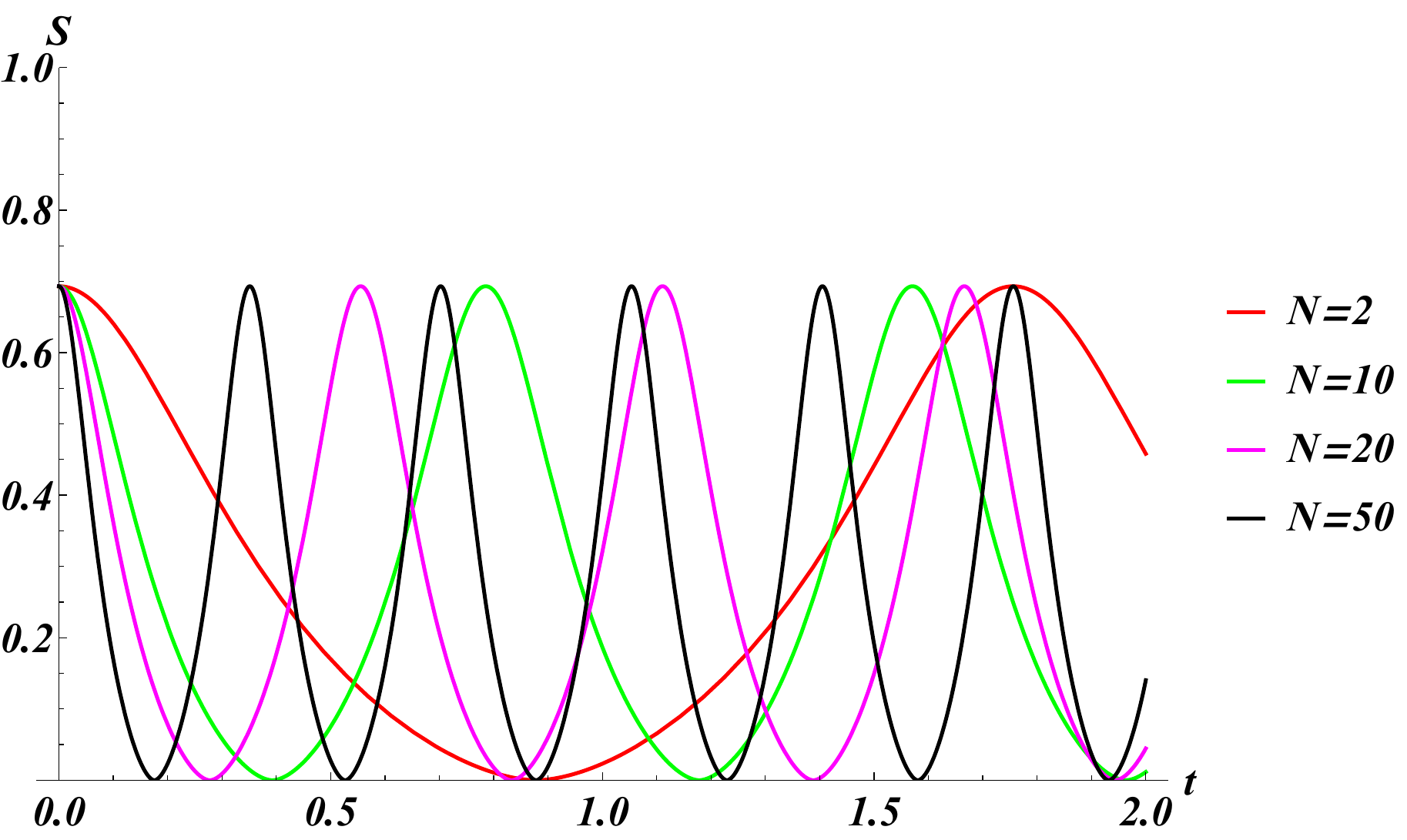}
	\caption{Von Neumann entropy as a function of time and varied bath size, with $c_1=1$, $g=0.7$, $\kappa=0.3$, $\kappa=1$}
	\label{fig:unbroken}
\end{figure}

Figure (\ref{fig:unbroken}) shows how the entropy evolves when $\kappa>g$. This is equivalent to the unbroken $PT$ regime of the non-Hermitian model. In this setting the entropy experiences so called "sudden death" similar to \cite{yonacc2006sudden}. The entropy rapidly decays from a maximum value to zero with a subsequent revival after the initial death. When the number of oscillators in the bath increases, the moment of vanishing entropy occurs at an earlier time.

\begin{figure}[H]
	\centering
	\includegraphics[scale=0.448]{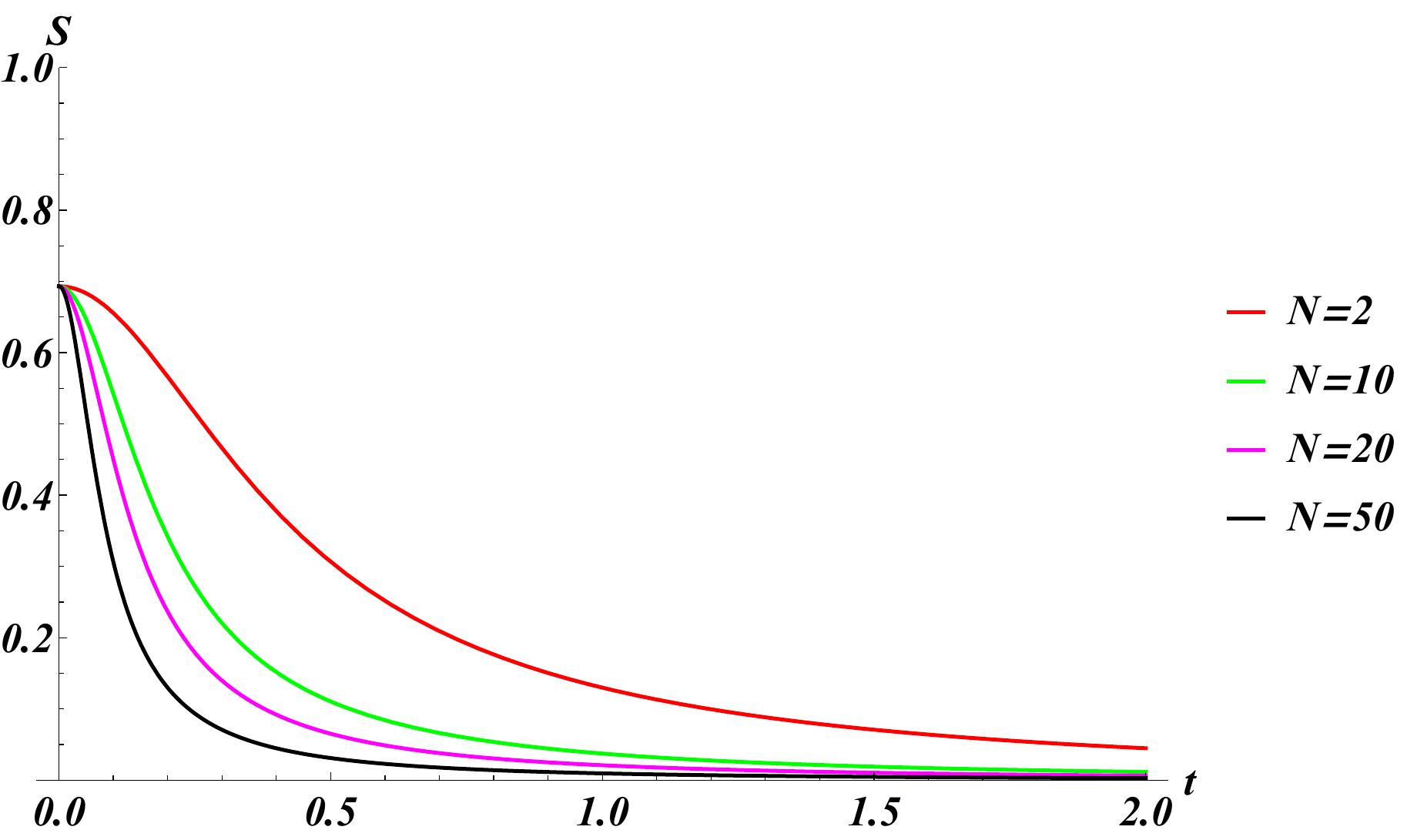}
	\caption{Von Neumann entropy as a function of time and varied bath size, with $c_1=1$, $g=\kappa$}
	\label{fig:exceptional}
\end{figure}

Figure (\ref{fig:exceptional}) depicts the entropy evolution when $\kappa=g$. This is equivalent to the exceptional point of the non-Hermitian model. In this specific setting, the system decays asymptotically from maximal entropy to zero. The half life of this decay decreases with the number of oscillators in the bath. 

\begin{figure}[H]
	\centering
	\includegraphics[scale=0.448]{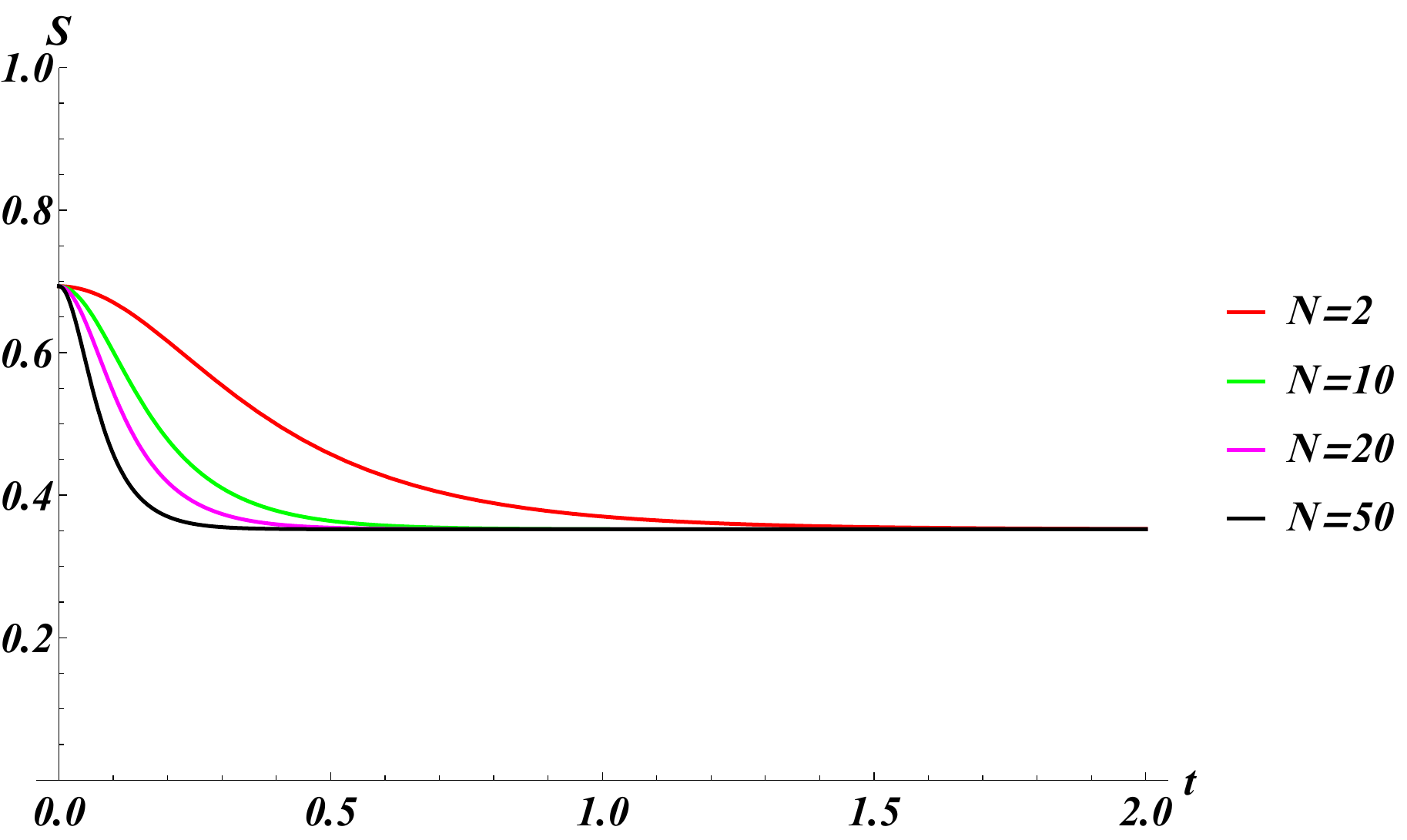}
	\caption{Von Neumann entropy as a function of time and varied bath size, with $c_1=1$, $g=0.3$, $\kappa=0.7$. The asymptote is at $S_{t\rightarrow\infty}\approx0.3521$}
	\label{fig:broken}
\end{figure}

Figure (\ref{fig:broken}) now shows the results of entropy evolution when $g>\kappa$. This is the spontaneously broken $PT$ regime of the original time-independent non-Hermitian model. In this case the system once again decays asymptotically but in this instance the decay is to a non-zero value of entropy. In this way, the entropy is preserved eternally. Once again the half life decreases with increasing $N$. The finite value that is asymptotically approached independently of $N$ is

\begin{equation}\label{min_entropy}
\begin{split}
\begin{aligned}
S_{t\rightarrow\infty}=&-\frac{1}{2}(1+\xi)\log\left[\frac{1}{2}(1+\xi)\right]\\
-&\frac{1}{2}(1-\xi)\log\left[\frac{1}{2}(1-\xi)\right],
\end{aligned}
\end{split}
\end{equation}
where 

\begin{equation}
\xi=\frac{\sqrt{c_1^2+g^2-\kappa^2}}{c_1}.
\end{equation}
We see the condition for the asymptote to exist is $|\frac{c1}{\sqrt{g^2-\kappa^2}}|>1$, which matches the reality condition of $\mu$ in equation (\ref{mu}).

We have found three significantly different phenomena at $\kappa>g$, $\kappa=g$ and $\kappa<g$. Specifically we see a change from rapid decay of entropy to zero, to asymptotic decay to zero through to asymptotic decay to a non-zero entropy. This can be interpreted as crossing the $PT$ exceptional point into the spontaneously broken regime of the original time-independent non-Hermitian system. However, with the existence of a time-dependent metric, the broken regime is no longer truly broken as we are able to provide a well-defined interpretation.

\section{Conclusion}

We derived a framework for the Von Neumann entropy in non-Hermitian quantum systems and applied it to a simple system bath coupled bosonic model. In order to analyse the model we were required to find a time-dependent metric and we chose to solve the time-dependent Dyson equation for this. This method also gave us the equivalent Hermitian system which we worked with to perform the analysis as the framework showed the entropy was equivalent in both systems. The $PT$ symmetry of the non-Hermitian system played an important role for the characterisation of the regimes of different qualitative behaviour in the evolution of the Von Neumann entropy. We found three different types of behaviour depending on whether we are in the $PT$ unbroken regime, at the exceptional point or in the spontaneously broken $PT$ regime. In the unbroken regime, the entropy underwent rapid decay to zero. At subsequent times it was revived and continued this oscillatory behaviour indefinitely. At the exceptional point, the entropy decayed asymptotically to zero and in the spontaneously broken regime, the entropy decayed asymptotically from a maximum to a finite minimum (\ref{min_entropy}) that remained constant with time. 

Our findings may have implications for maintaining entanglement in quantum computers when the computer is operated in the spontaneously broken $PT$ regime. The challenge here is to construct a system in a laboratory that mimics that of the non-Hermitian system presented here. However, non-Hermitian systems have been realised in quantum optical experiments \cite{guo2009observation,ruter2010observation} and so it is certainly possible that the same could be carried in quantum computing. \\\\
\noindent\textbf{ACKNOWLEDGEMENTS}: TF is supported by a City, University of London Research Fellowship.

\bibliographystyle{frithstyle}
\bibliography{bibliography}

\begin{thebibliography}{10}
\expandafter\ifx\csname urlstyle\endcsname\relax
  \providecommand{\doi}[1]{doi:\discretionary{}{}{}#1}\else
  \providecommand{\doi}{doi:\discretionary{}{}{}\begingroup
  \urlstyle{rm}\Url}\fi

\bibitem{nielsen2002quantum}
M.~A. Nielsen and I.~Chuang.
\newblock Quantum computation and quantum information (2002).

\bibitem{bennett2000quantum}
C.~H. Bennett and D.~P. DiVincenzo.
\newblock Quantum information and computation.
\newblock \emph{Nature} 404(6775), 247 (2000).

\bibitem{raussendorf2001one}
R.~Raussendorf and H.~J. Briegel.
\newblock A one-way quantum computer.
\newblock \emph{Phys. Rev. Lett.} 86(22), 5188 (2001).

\bibitem{steane1998quantum}
A.~Steane.
\newblock Quantum computing.
\newblock \emph{Rep. Prog. Phys} 61(2), 117 (1998).

\bibitem{yu2009sudden}
T.~Yu and J.~Eberly.
\newblock Sudden death of entanglement.
\newblock \emph{Science} 323(5914), 598--601 (2009).

\bibitem{yonacc2006sudden}
M.~Y{\"o}na{\c{c}}, T.~Yu and J.~Eberly.
\newblock Sudden death of entanglement of two Jaynes--Cummings atoms.
\newblock \emph{J. Phys. B: Atomic, Molecular and Optical Physics} 39(15), S621
  (2006).

\bibitem{unruh1995maintaining}
W.~G. Unruh.
\newblock Maintaining coherence in quantum computers.
\newblock \emph{Phys. Rev. A} 51(2), 992 (1995).

\bibitem{palma1996quantum}
G.~M. Palma, K.-A. Suominen and A.~Ekert.
\newblock Quantum computers and dissipation.
\newblock \emph{Proc. Royal Soc. Lond. A: Math, Phys and Eng} 452(1946),
  567--584 (1996).

\bibitem{scolarici2006time}
G.~Scolarici and L.~Solombrino.
\newblock Time evolution of non-Hermitian quantum systems and generalized
  master equations.
\newblock \emph{Czechoslovak J. Phys} 56(9), 935--941 (2006).

\bibitem{croke2015pt}
S.~Croke.
\newblock PT-symmetric Hamiltonians and their application in quantum
  information.
\newblock \emph{Phys. Rev. A} 91(5), 052,113 (2015).

\bibitem{kawabata2017information}
K.~Kawabata, Y.~Ashida and M.~Ueda.
\newblock Information retrieval and criticality in parity-time-symmetric
  systems.
\newblock \emph{Phys. Rev. Lett.} 119(19), 190,401 (2017).

\bibitem{jones2010quantum}
H.~Jones and E.~Moreira~Jr.
\newblock Quantum and classical statistical mechanics of a class of
  non-Hermitian Hamiltonians.
\newblock \emph{J. Phys. A: Math. Theor.} 43(5), 055,307 (2010).

\bibitem{bender1998real}
C.~M. Bender and S.~Boettcher.
\newblock Real spectra in non-Hermitian Hamiltonians having $\mathcal{PT}$
  symmetry.
\newblock \emph{Phys. Rev. Lett.} 80(24), 5243 (1998).

\bibitem{bender2007making}
C.~M. Bender.
\newblock Making sense of non-Hermitian Hamiltonians.
\newblock \emph{Rep. Prog. Phys.} 70(6), 947 (2007).

\bibitem{mostafazadeh2010pseudo}
A.~Mostafazadeh.
\newblock Pseudo-Hermitian representation of quantum mechanics.
\newblock \emph{Int. J. Geom. Methods Mod. Phys.} 7(7), 1191--1306 (2010).

\bibitem{moiseyev2011non}
N.~Moiseyev.
\newblock Non-Hermitian quantum mechanics.
\newblock Cambridge University Press (2011).

\bibitem{scholtz1992quasi}
F.~Scholtz, H.~Geyer and F.~Hahne.
\newblock Quasi-Hermitian operators in quantum mechanics and the variational
  principle.
\newblock \emph{Ann. Phys. (N. Y.)} 213(1), 74--101 (1992).

\bibitem{znojil2006construction}
M.~Znojil and H.~B. Geyer.
\newblock Construction of a unique metric in quasi-Hermitian quantum mechanics:
  nonexistence of the charge operator in a 2$\times$ 2 matrix model.
\newblock \emph{Phys. Lett. B} 640(1-2), 52--56 (2006).

\bibitem{siegl2012metric}
P.~Siegl and D.~Krej{\v{c}}i{\v{r}}{\'\i}k.
\newblock On the metric operator for the imaginary cubic oscillator.
\newblock \emph{Phys. Rev. D} 86(12), 121,702 (2012).

\bibitem{castro2009spin}
O.~A. Castro-Alvaredo and A.~Fring.
\newblock A spin chain model with non-Hermitian interaction: the Ising quantum
  spin chain in an imaginary field.
\newblock \emph{J. Phys. A: Math. and Theor} 42(46), 465,211 (2009).

\bibitem{jones2007scattering}
H.~Jones.
\newblock Scattering from localized non-Hermitian potentials.
\newblock \emph{Phys. Rev. D} 76(12), 125,003 (2007).

\bibitem{musumbu2006choice}
D.~Musumbu, H.~Geyer and W.~Heiss.
\newblock Choice of a metric for the non-Hermitian oscillator.
\newblock \emph{J. Phys A: Math and Theor} 40(2), F75 (2006).

\bibitem{mostafazadeh2006metric}
A.~Mostafazadeh.
\newblock Metric operator in pseudo-Hermitian quantum mechanics and the
  imaginary cubic potential.
\newblock \emph{J. Phys. A: Math and Gen} 39(32), 10,171 (2006).

\bibitem{mostafazadeh2007time}
A.~Mostafazadeh.
\newblock Time-dependent pseudo-Hermitian Hamiltonians defining a unitary
  quantum system and uniqueness of the metric operator.
\newblock \emph{Phys. Lett. B} 650(2-3), 208--212 (2007).

\bibitem{fring2016non}
A.~Fring and M.~H. Moussa.
\newblock Non-Hermitian Swanson model with a time-dependent metric.
\newblock \emph{Phy. Rev. A} 94(4), 042,128 (2016).

\bibitem{fring2016unitary}
A.~Fring and M.~H. Moussa.
\newblock Unitary quantum evolution for time-dependent quasi-Hermitian systems
  with nonobservable Hamiltonians.
\newblock \emph{Phys. Rev. A} 93(4), 042,114 (2016).

\bibitem{maamache2017pseudo}
M.~Maamache, O.~K. Djeghiour, N.~Mana and W.~Koussa.
\newblock Pseudo-invariants theory and real phases for systems with
  non-Hermitian time-dependent Hamiltonians.
\newblock \emph{Eur. Phys. J} 132(9), 383 (2017).

\bibitem{znojil2008time}
M.~Znojil.
\newblock Time-dependent version of crypto-Hermitian quantum theory.
\newblock \emph{Phy. Rev. D} 78(8), 085,003 (2008).

\bibitem{de2006time}
C.~F. de~Morisson~Faria and A.~Fring.
\newblock Time evolution of non-Hermitian Hamiltonian systems.
\newblock \emph{J. Phys. A: Math. Gen.} 39(29), 9269 (2006).

\bibitem{mostafazadeh2018energy}
A.~Mostafazadeh.
\newblock Energy observable for a quantum system with a dynamical Hilbert space
  and a global geometric extension of quantum theory.
\newblock \emph{Phys. Rev. D} 98(4), 046,022 (2018).

\bibitem{ExactSols}
A.~Fring and T.~Frith.
\newblock Exact analytical solutions for time-dependent {H}ermitian
  {H}amiltonian systems from static unobservable non-{H}ermitian
  {H}amiltonians.
\newblock \emph{Phys. Rev. A} 95(1), 010,102 (2017).

\bibitem{FRING20172318}
A.~Fring and T.~Frith.
\newblock Mending the broken $\mathcal{PT}$-regime via an explicit
  time-dependent {D}yson map.
\newblock \emph{Phys. Lett. A} 381(29), 2318 -- 2323 (2017).

\bibitem{HigherSpin}
A.~Fring and T.~Frith.
\newblock Metric versus observable operator representation, higher spin models.
\newblock \emph{Eur. Phys. J} 133(2), 57 (2018).

\bibitem{fring2018tdm}
A.~Fring and T.~Frith.
\newblock Time-dependent metric for the two dimensional, non-Hermitian coupled
  oscillator.
\newblock \emph{arXiv preprint arXiv:1812.02862}  (2018).

\bibitem{HigherDims}
A.~Fring and T.~Frith.
\newblock Solvable two-dimensional time-dependent non-Hermitian quantum systems
  with infinite dimensional Hilbert space in the broken $\mathcal{PT}$-regime.
\newblock \emph{J. Phys. A: Math. Theor.} 51(26), 265,301 (2018).

\bibitem{guo2009observation}
A.~Guo, G.~Salamo, D.~Duchesne, R.~Morandotti, M.~Volatier-Ravat, V.~Aimez,
  G.~Siviloglou and D.~Christodoulides.
\newblock Observation of $\mathcal{PT}$-symmetry breaking in complex optical
  potentials.
\newblock \emph{Phys. Rev. Lett.} 103(9), 093,902 (2009).

\bibitem{ruter2010observation}
C.~E. R{\"u}ter, K.~G. Makris, R.~El-Ganainy, D.~N. Christodoulides, M.~Segev
  and D.~Kip.
\newblock Observation of parity--time symmetry in optics.
\newblock \emph{Nat. Phys.} 6(3), 192 (2010).

\end{thebibliography}

\end{multicols}

\end{document}